\documentclass[cits]{PoS}

\title{The MAGIC telescope}

\ShortTitle{The MAGIC telescope}

\author{\speaker{Ciro Bigongiari for the MAGIC collaboration}
        \thanks{Updated MAGIC members list at http://wwwmagic.mppmu.mpg.de/collaboration/index.html}\\
        Padua University and INFN Padua, Italy\\
        E-mail: \email{ciro.bigongiari@pd.infn.it}}

\abstract{
MAGIC (Major Atmospheric Gamma Imaging Cherenkov telescope) is presently the largest ground-based gamma ray telescope. MAGIC has been taking data regularly since October 2004 at the Roque de los Muchachos Observatory on the island of La Palma. 
In this paper the MAGIC telescope status, its performances and some preliminary results on observed gamma ray 
sources are presented. 
}

\FullConference{International Europhysics Conference on High Energy Physics\\
		 July 21st - 27th 2005\\
		 Lisboa, Portugal}
 
\begin{document}

\section{Introduction}

MAGIC (Major Atmospheric Gamma Imaging Cherenkov) telescope 
\cite{Barrio_1988}, 
is presently the largest imaging air Cherenkov telescope in operation. 
MAGIC has a 17~m diameter, $f/D = 1$, parabolic reflector covering a total surface of $234~{\mathrm m}^2$.
The reflector dish is composed of 956 
($0.495 \times 0.495~{\mathrm m}^2$) 
diamond milled aluminum mirrors  
\cite{Bigongiari_2004}. 
The reflector shape is parabolic to minimize the time spread of Cherenkov light flashes on the focal plane. 
Aluminum mirrors were chosen instead of glass ones to reduce the weight of the reflecting surface and allow a fast slewing of the telescope. For the same reason the telescope frame is made of carbon fiber tubes. A very short slewing time is needed to catch the $\gamma$-ray prompt emission by GRBs. 
The reflected Cherenkov photons are recorded by a 
$3.5^{\circ}-3.8^\circ$ 
FOV hexagonal camera in the telescope focal plane, composed by 397 
$0.1^\circ$ 
FOV photomultiplier tubes, surrounded by 180 
$0.2^\circ$ 
FOV PMTs. 
The PMTs have hemispherical windows and only 6 dynodes to minimize the time response width. The PMT photo conversion efficiency has been enhanced up to 30\% and extended to the UV by coating the window with wavelength shifter 
\cite{Paneque_2004}. 
The PMT signals are transferred via optical fibers 
\cite{Paneque_2003} 
to the electronic room where they are split and sent to trigger and digitizing systems. 
The trigger decision is generated by a 2-level system using the signals of the 325 innermost PMTs. Only signals above an adjustable threshold are considered. The time coincidence within 6~nS of signals from 4 adjacent PMTs is required
\cite{Meucci_2004}. 
The analog signals are continuously digitized by 8 bit 300~MHz Flash ADCs. 
If the trigger condition is fulfilled the signals stored in FADC ring buffers are 
written to a FIFO buffer and saved by the DAQ.
 
\section{MAGIC performances} 

The MAGIC construction was completed in Fall 2003 at the Roque de Los Muchachos observatory 
on the Canary island of La Palma, 
($28.75^{\circ}$ N, $17.90^{\circ}$ W, 2200~m above sea level). 
The commissioning phase took one year and 
in Fall 2004 MAGIC started to take data continuously on $\gamma$-ray sources. The first regular 
observation cycle started in April 2005. 
All the technical innovations implemented in the MAGIC construction 
\cite{Mirzoyan_2003} 
are now working without major problems and most of the telescope parameters are well within the design specifications
\cite{Cortina_2005}.
According to detailed simulation of atmospheric showers and detector response  
the trigger threshold is around 60~GeV
\cite{Majumdar_2005}. 
Presently the analysis threshold is about $E_{Th} = 100$~GeV as the $\gamma$/hadron discrimination becomes more and more difficult at lower energies. There is still room for lowering both trigger and analysis thresholds with present hardware fully exploiting second level trigger capabilities and improving the analysis technique. 
MAGIC integral flux sensitivity has been calculated from MC simulation and results in 
%
about 5\% of ${\Phi}_{Crab}$ at $E > 100$~GeV and 2\% of ${\Phi}_{Crab}$ at $E > 1$~TeV
\cite{Majumdar_2005}. 
The angular resolution has been estimated applying the DISP method to Crab data
and results in about $0.1^{\circ}$ for $\gamma$-ray events with $E > 200$~GeV  \cite{Domingo_2005}.
The accuracy in point-source position determination improves as the square root of the number of collected events 
and is ultimately limited by tracking accuracy ($\simeq 0.01^{\circ}$)
\cite{Riegel_2005}.
The energy resolution has been estimated from MC data and results in ${\Delta}E/E \simeq 30\%$ at $E = 100$~GeV and 
${\Delta}E/E \simeq 20\%$ for $E > 1$~TeV 
\cite{Wagner_2005}.  

\section{Gamma-ray source observations} 

Until Nov 2005 MAGIC has observed nearly 50 $\gamma$-ray source candidates. Emission of $\gamma$-rays has been detected by 8 of them, 4 galactic and 4 extragalactic. 
\par
The first and most observed source is the Crab Nebula. 
The Crab spectrum between some tens of GeV and some hundreds of GeV is  
astrophysicaly very interesting because the inverse Compton peak is expected to be close to 100~GeV 
and the cut-off of the pulsed $\gamma$-emission between 10 and 100~GeV.  
MAGIC has measured the Crab spectrum 
down to 100~GeV 
\cite{Wagner_2005} 
for the first time, 
definitely heading for the inverse Compton peak.   
\begin{figure}[thb]
 \begin{minipage}[b]{0.45\textwidth}
   \centering
    \includegraphics[width=0.9\textwidth]{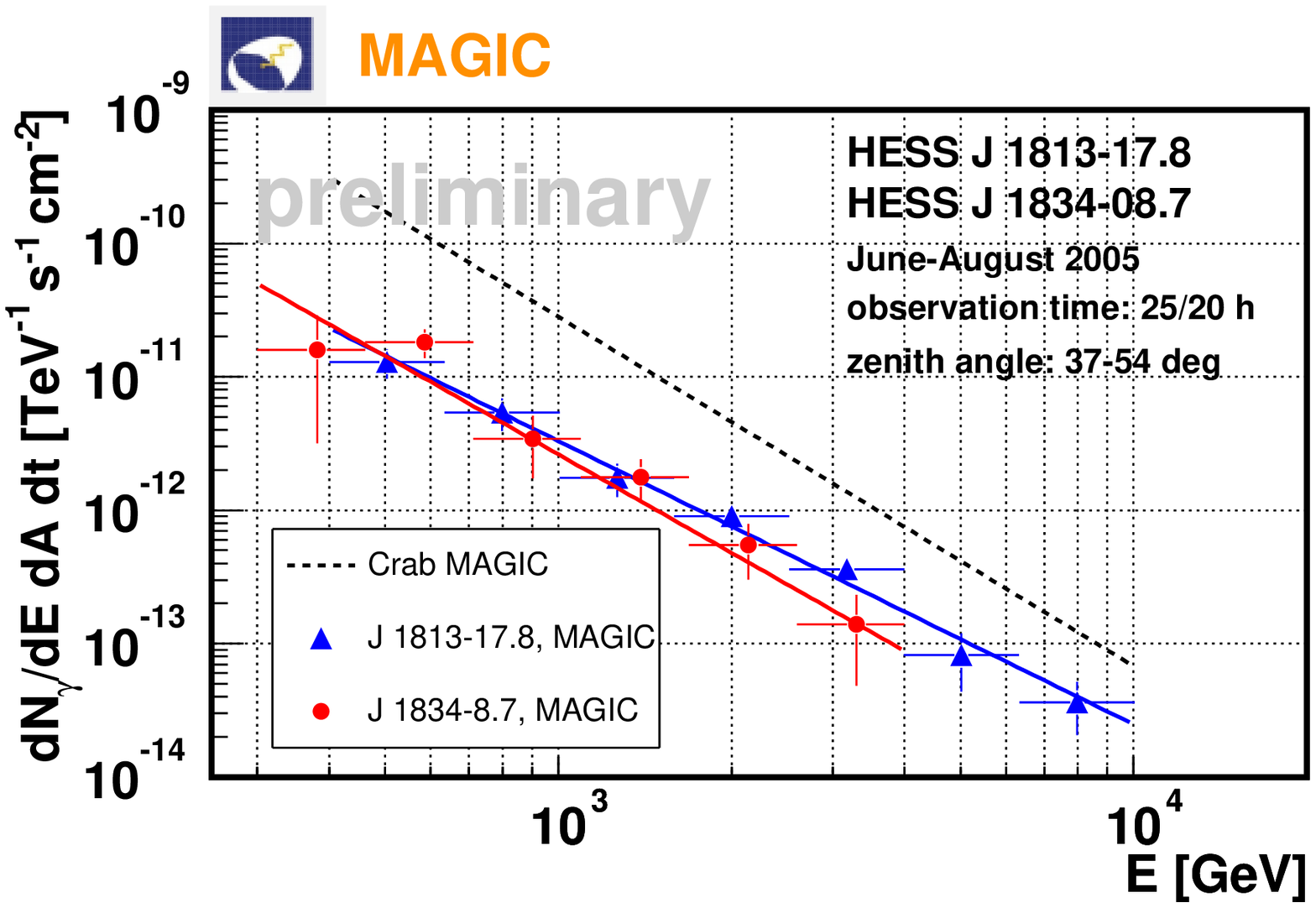}
     \caption{Observed differential spectra of HESS~J1813-178 and HESS~J1834-087.} 
   \label{1813_spectrum}
 \end{minipage}
\ \hspace{5mm} \
 \begin{minipage}[b]{0.45\textwidth}
  \centering
    \includegraphics[width=0.9\textwidth]{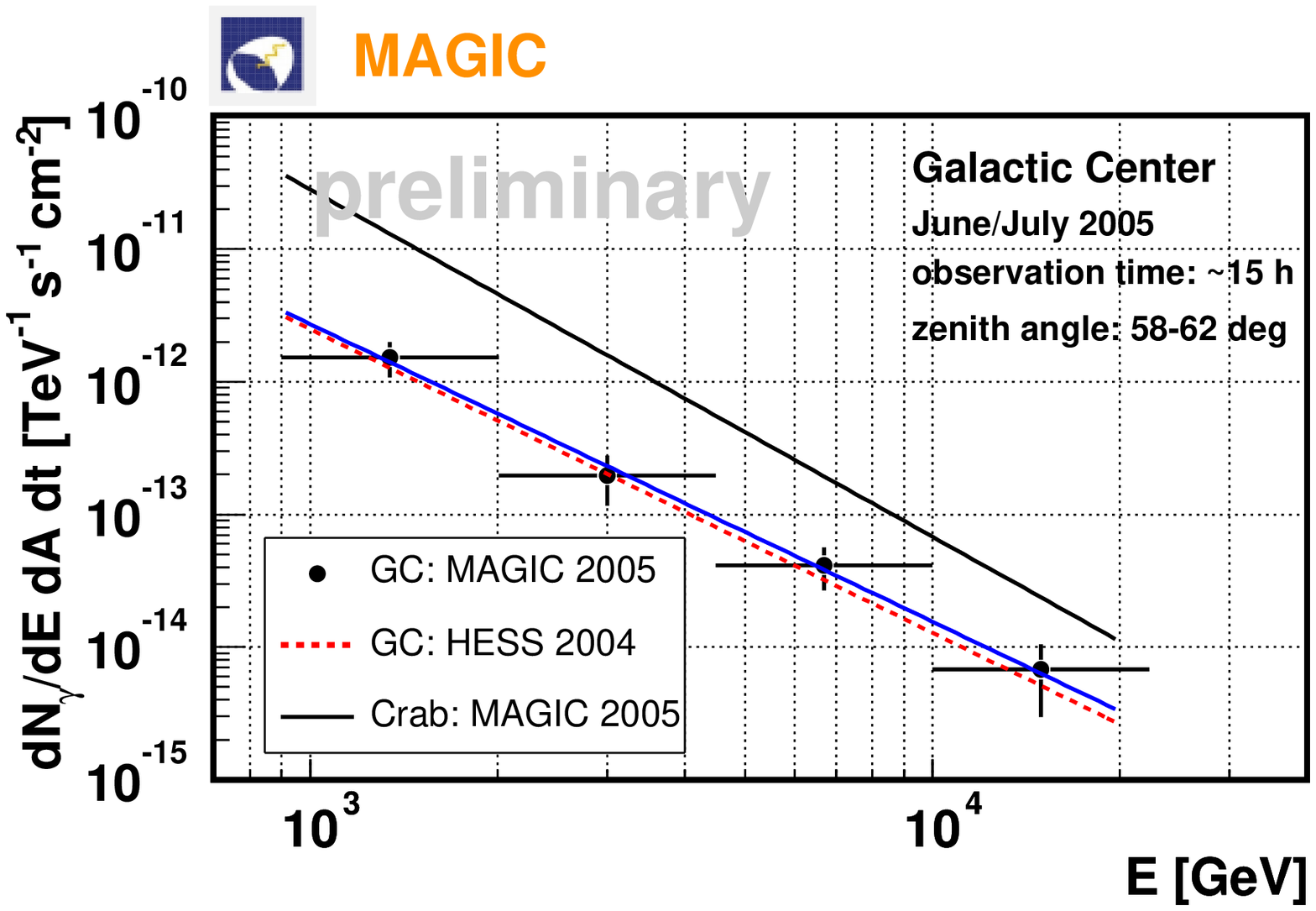}
     \caption{Observed differential spectrum of the source at galactic center.} 
   \label{GC_spectrum}
 \end{minipage}
\end{figure}
MAGIC detected $\gamma$-ray emission from the SNRs, HESS~J1813-178 and HESS~J1834-087 \cite{Magic_1813_1834}, 
recently discovered by HESS \cite{HESS_1813_1834}. MAGIC can observe these sources only at zenith angle around $40^{\circ}-50^{\circ}$. This implies a higher energy threshold $E_{Th} = 300$~GeV but also a larger effective area. 
The spectra measured by MAGIC are in good agreement with HESS ones (Figure \ref{1813_spectrum}).  
The emission of VHE $\gamma$-rays from a source at galactic center discovered by Cangaroo 
and recently confirmed by HESS 
has been observed by MAGIC too \cite{Magic_GC}. Within errors the flux measured by MAGIC is compatible with the hard spectrum measured by HESS (Figure \ref{GC_spectrum}). The source position measured by MAGIC is compatible with both Sgr~$A^{*}$ and Sgr~A~East. 
\par
MAGIC detected $\gamma$-rays from three known extragalactic sources, Mrk~421, Mrk~501 and 1ES~1959+650 and discovered 
$\gamma$-ray emission from another AGN, 1ES~1218+304, which is currently the second most distant gamma-ray source.
Mrk~421 spectrum has been measured down to 100~GeV \cite{Mazin_2005} for the first time. 
Flux variations of a factor 2 on an hourly time scale were observed with no evidence of spectral index variations.   
MAGIC detected an intense flare ($\Phi \simeq 4 \dot {\Phi}_{Crab}$) from Mrk~501 on June $30^{th}$,  2005, 
while monitoring this source and alerted the astronomical community (IAU circular 8562). 
Very large flux variations were observed on short time scale. There are hints of spectral hardening when the 
source is in high flux state.
The AGN 1ES~1959+650 spectrum has been measured by MAGIC down to 200~GeV when it was in low flux state 
\cite{Magic_1959}. 
This observation is characterized by lack of strong time variability and the presence of a significant VHE 
$\gamma$-ray emission in the absence of strong X-ray or optical emission. 
MAGIC discovered a new $\gamma$-ray source, the AGN 1ES~1218+304, and measured its spectrum between 100~GeV and 1~TeV. 
The measurement of VHE $\gamma$-ray spectra of distant AGNs, like 1ES~1218+304 can be used to evaluate the Metagalactic Radiation Field. 
 
\section{Conclusions and outlook}

MAGIC has been taking data regularly since Fall 2004. About 50 possible $\gamma$-ray sources have been observed so far 
and $\gamma$-ray emission has been detected by 8 of them. MAGIC confirmed $\gamma$-ray emission from two sources 
recently discovered by HESS and detected $\gamma$-rays from 1ES~1218+304 at $z = 0.182$ for the first time. 
MAGIC is able to measure the emission spectrum of sources which culminate close to zenith down to 100~GeV. This limit will be lowered further by improving the data analysis technique, adopting new 2.5~GHz FADCs and building a second telescope.    
The construction of the second telescope, already started, will be completed by the end of 2007. 




\end{document}